\documentstyle[aps,twocolumn,epsfig,floats]{revtex}

\draft
\begin{document}

\twocolumn[\hsize\textwidth\columnwidth\hsize\csname @twocolumnfalse\endcsname

\title{ 
Electronic structure of metallic antiperovskite compound GaCMn$_3$
}

\author{ J. H. Shim, S. K. Kwon, and B. I. Min}

\address{ Department of Physics, Pohang University of Science and
Technology, Pohang 790-784, Korea}

\date{\today}
\maketitle

\begin{abstract}
We have investigated electronic structures of antiperovskite
GaCMn$_3$ and related Mn compounds SnCMn$_3$, ZnCMn$_3$, and ZnNMn$_3$.
In the paramagnetic state of GaCMn$_3$, the Fermi
surface nesting feature along the $\Gamma{\rm R}$ direction is observed,
which induces the antiferromagnetic (AFM) spin ordering 
with the nesting vector {\bf Q} $\sim \Gamma{\rm R}$.
Calculated susceptibilities confirm the nesting scenario 
for GaCMn$_3$ and also explain various magnetic structures of 
other antiperovskite compounds. Through the band folding effect, 
the AFM phase of GaCMn$_3$ is stabilized.
Nearly equal densities of states at the Fermi level
in the ferromagnetic and AFM phases of GaCMn$_3$
indicate that two phases are competing in the ground state.
\end{abstract}

]

%
Due to recent observations of giant-magnetoresistance (GMR)
phenomenon in GaCMn$_3$\cite{kamishima00,kim} and superconductivity
in MgCNi$_3$\cite{he}, compounds with antiperovskite structure
have attracted great attention.
Especially, antiperovskite manganese compounds that exhibit
intriguing magnetic structures have been studied 
intensively\cite{fruchart70,fruchart78,kaneko,fruchart73}.
Although the antiperovskite structure has an analogy to the 
perovskite oxides, their physical properties are very different.
The antiperovskite structure locates transition metal 
elements at the octahedral corners, not at the centers.  
For example, in the antiperovskite GaCMn$_3$, C is centered in
the octahedron composed of six Mn atoms, and Mn has only two nearest 
neighbor C atoms in contrast to the perovskite oxides in which 
a transition metal has six nearest neighbor oxygens.

GaCMn$_3$ exhibiting the GMR phenomenon has the
antiferromagnetic (AFM) phase in the ground state\cite{fruchart70}.
Spins located on the (111) plane are ferromagnetically coupled and 
polarized collinearly along the [111] direction. 
The spin directions in these ferromagnetic (FM) planes are
alternating along the [111] direction to have the AFM phase.
Upon heating, it shows a first order transition to the FM phase 
at the transition temperature $T_t \sim 160\ {\rm K}$
through a small region of the intermediate state where
both the AFM and FM components are present\cite{kamishima00,kim}.
$T_t$ decreases with increasing magnetic field or pressure\cite{kamishima98}.
The transition is accompanied by the discontinuous suppression of 
volume and resistivity.  The GMR phenomenon is manifested in this region.
$T_t$ also depends on the C stoichiometry. That is,
with small deficit of C ($\sim$ 4\%),
$T_t$ is reduced to 0 K and the AFM region disappears\cite{fruchart78}.
GaCMn$_3$ has the Curie temperature of $T_{\rm C} \sim 250\ {\rm K}$.
The magnetic moment per Mn is $\mu_{\rm Mn} = 1.8\pm 0.1\ \mu_{\rm B}$ 
at 4.2 K in the AFM phase, and $\mu_{\rm Mn} = 1.3\pm 0.1\ \mu_{\rm B}$ 
at 193 K in the FM phase\cite{fruchart78}. 
On the other hand, the recent neutron diffraction study shows that 
$\mu_{\rm Mn} \sim 1\ \mu_{\rm B}$ for both phases\cite{kim}.
The relatively small magnetic moment, as compared to 
other manganese intermetallic compounds, reflects that this system would 
be described by the itinerant magnetism. 

The spin configuration and the magnetic structure vary in a complicated
way by electron or hole doping in the Mn antiperovskite system\cite{kaneko}.
For effectively one-hole doped ZnCMn$_3$, the spin configuration becomes 
non-collinear with the tetragonal magnetic structure,
and the AFM and FM phases coexist in the ground state\cite{fruchart73}.
This coexisting phase is similar in nature to the intermediate phase 
observed in GaCMn$_3$\cite{kamishima00}.
There is a first order transition to the FM
phase with $T_{\rm C} \sim 350\ {\rm K}$.
Carbon deficient GaC$_{0.935}$Mn$_3$ also has a similar magnetic structure 
to ZnCMn$_3$. 
For effectively one-electron doped SnCMn$_3$,
the spin configuration is also non-collinear up to
$T_{\rm C} \sim 290\ {\rm K}$ with a complicated magnetic structure.
Differently from the above compounds, SnCMn$_3$ does not have
a stable FM phase.
On the other hand, ZnNMn$_3$ which has the same number of valence electrons
as GaCMn$_3$ shows different magnetic structure and properties\cite{kim-new}.
It also has an AFM phase in the ground state, but no stable FM phase at
higher temperature. Upon heating, it shows a direct transition from the
AFM to the paramagnetic (PM) phase.  
Moreover, the AFM phase itself shows a different magnetic 
configuration. Spins on the (111) plane are non-collinear
with clockwise or counterclockwise spin current configurations,
and so the spin moments in the plane are compensating each other.
Spin current directions in such (111) planes, that is, the spin chiralities
are alternating along the [111] direction\cite{fruchart78}.
The origin of such intriguing magnetic structures in antiperovskite 
Mn compounds has not been addressed yet.

In this paper, we have studied electronic structures of GaCMn$_3$
and related antiperovskite Mn compounds such as ZnCMn$_3$, SnCMn$_3$, and 
ZnNMn$_3$. Band structures in their PM phases indicate that 
the magnetic structures are strongly correlated with the Fermi surface 
nesting. To investigate the correlation 
between the magnetic structure and the Fermi surface nesting, 
we have evaluated the susceptibility, and
discussed the doping effects on the susceptibility and the
magnetic structure. 
We have also studied the electronic structures of
both AFM and FM phases of GaCMn$_3$ and compared them with that of
the PM phase.

%
We have used the linearized muffin-tin orbital (LMTO) band method in 
the local spin density approximation (LSDA). 
We include the valence band muffin-tin
orbitals up to $d$ for Ga and Mn, $p$ for C. Ga 3$d$ state is 
considered as core state. We consider the cubic structure with lattice
constant $a$ = 3.89 ${\rm\AA}$ for all magnetic structures of GaCMn$_3$.
In reality, the AFM phase of GaCMn$_3$ has slightly larger lattice constant 
than the FM phase by about 0.1 \% near the transition temperature \cite{kim}.
However, such a small difference in the lattice constant gives minor effect
on the electronic structure.
For the calculation of AFM structure, we have considered a doubled 
unit cell along the [111] direction which is trigonal. Atomic sphere 
radii of Ga, C, and Mn are employed as 1.5, 0.8, 1.4 ${\rm\AA}$, respectively.
To evaluate susceptibilities, we have also obtained electronic
structures of related compounds, SnCMn$_3$, ZnCMn$_3$, and ZnNMn$_3$,
with lattice constants 3.99, 3.93, 3.89 ${\rm\AA}$,
respectively\cite{kaneko,kim-new}.
Because Zn 3$d$ state is located near $E_{\rm F}$, it is treated 
as valence band.
The other structural parameters are the same as in the case of GaCMn$_3$.
\begin{figure}[hb]
\epsfig{file=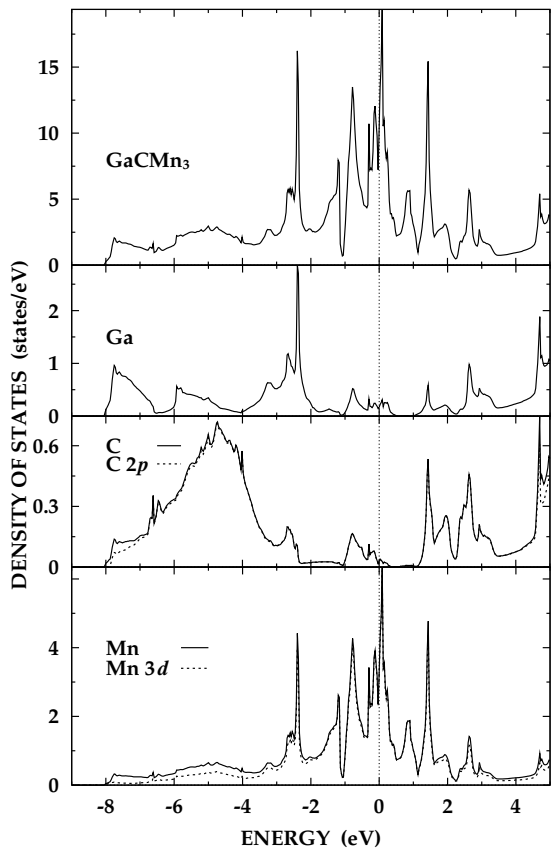,width=7.5cm}
\caption{The total and atomic site projected local DOSs of GaCMn$_3$
in the PM phase. The DOS peak near $-12$ eV which is mainly 
composed of C 2$s$ state is not shown.}
\label{dos_para}
\end{figure}
\begin{figure}[htb]
\epsfig{file=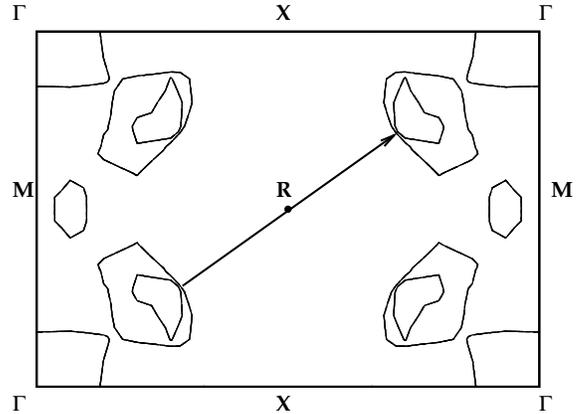,width=7.5cm}
\caption{The Fermi surface of GaCMn$_3$ on the (110) plane of the 
simple-cubic BZ. 
The arrow corresponds to the nesting vector {\bf Q} along the [111] direction.}
\label{fer}
\end{figure}
%

%
Figure \ref{dos_para} shows the total density of states (DOS)
and projected local density of states (PLDOS) 
of GaCMn$_3$ in its PM phase.
There exists strong hybridization between Mn 3$d$ and two neighboring 
C 2$p$ states, which gives rise to the extended bonding states between
$-3$ eV and $-7$eV.  Due to this hybridization, the band width of Mn 3$d$
states becomes wide so that the system behaves as an itinerant 
electron system.  At the Fermi level $E_{\rm F}$, 
Mn 3$d$ nonbonding states are located, and so the DOS at $E_{\rm F}$
is mainly due to Mn 3$d$ states.

Band structure of GaCMn$_3$ indicates that three bands cross the Fermi level, 
forming the Fermi surfaces roughly consistent with Ref.\cite{motizuki}. 
Figure \ref{fer} provides the Fermi surfaces in the (110) plane
of the simple-cubic Brillouin zone (BZ).
Noteworthy is the nesting feature along the [111] direction,
which is expected to be closely related to the AFM ground state of GaCMn$_3$. 
The size of the nesting vector is ${\bf Q} = \frac{9}{10}\Gamma{\rm R}$. 
It thus reflects that 
the Fermi surface nesting would induce the spin ordering along
the [111] direction with wave vector ${\bf Q} \sim \Gamma {\rm R}$, 
resulting in the AFM structure observed in GaCMn$_3$.

In order to examine the nesting feature more clearly,
we have evaluated the susceptibility which would drive 
the magnetic instability.
The bare electronic susceptibility $\chi_0$ can be obtained using the
band structure output\cite{myron}:
\begin{equation}
      \chi_0({\rm\bf q}) = \frac{1}{N}\sum_{n,n',{\rm\bf k}} 
      \frac{f(\epsilon_{n,{\rm\bf k}})[1-f(\epsilon_{n',{\rm\bf k+q}})]}
      {\epsilon_{n',{\rm\bf k+q}} - \epsilon_{n,{\rm\bf k}}},
\end{equation}
where $f(\epsilon$) is the Fermi-Dirac distribution function,
$\epsilon_{n,{\rm\bf k}}$ and $\epsilon_{n',{\rm\bf k+q}}$ are the 
eigenvalues at {\bf k} and {\bf k+q} of the first BZ with 
the band indices $n$ and $n'$.
Some ordering such as charge or spin density wave is induced with
wave vector {\bf Q} which maximizes $\chi_0({\bf Q})$.
In this system, we have considered only three bands crossing the Fermi
level to obtain the susceptibility.

\begin{figure}[tb]
\epsfig{file=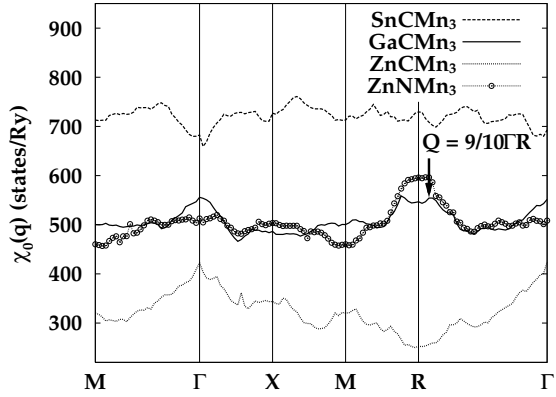,angle=270,width=7.5cm}
\caption{The calculated susceptibilities of GaCMn$_3$ and related compounds.
The vertical arrow denotes the local maximum of the susceptibility in GaCMn$_3$
with wave vector ${\rm\bf Q} = \frac{9}{10}\Gamma{\rm R}$.
SnCMn$_3$ and ZnCMn$_3$ are effectively one electron and hole doped 
compounds respectively, and ZnNMn$_3$ has the same number of
valence electrons as GaCMn$_3$.
}
\label{suscep}
\end{figure}
Figure \ref{suscep} shows the calculated susceptibilities of GaCMn$_3$ and 
related compounds, SnCMn$_3$, ZnCMn$_3$, and ZnNMn$_3$.
In the susceptibility of GaCMn$_3$, local maxima are observed at point
$\Gamma$ and near point R. This feature suggests that the FM phase with
${\bf Q}=0$ or the AFM phase with ${\bf Q}\sim \Gamma{\rm R}$ would be stabilized
depending on the situation. The {\bf k} point denoted by vertical arrow 
in Fig. \ref{suscep} corresponds to the nesting vector 
{\bf Q}($=\frac{9}{10}\Gamma{\rm R}$). Similar magnitudes of
$\chi_0$ at points $\Gamma$ and R indicates that both phases are competing.

For one-electron doped SnCMn$_3$, the susceptibility at point $\Gamma$ is
much suppressed, suggesting that the FM phase is unstable.
This feature is consistent with the experiment.
In the case of one-hole doped ZnCMn$_3$, the susceptibility at point R is 
suppressed and local maxima are observed at other {\bf k} points including
$\Gamma$.
It would assist the appearance of complicated magnetic structures other than 
the AFM structure in the ground state.
As for ZnNMn$_3$, the substantial enhancement of $\chi_0$ is observed
at point R, whereas $\chi_0$ at point $\Gamma$ is suppressed. 
This feature is again consistent with the experimental result
that the AFM phase is stabilized upon cooling without 
going through the FM phase.
\begin{table}[b]
\setdec 0.00
\caption{The calculated total and partial DOSs at $E_{\rm F}$ (in states/eV)
and magnetic moments  (in $\mu_{\rm B}$/f.u.)}
\begin{tabular}{lccccccccc}
 &$ N_{\rm total}$  &$ N_{\rm Ga}$ 
 &$ N_{\rm C}$  &$ N_{\rm Mn}$
 &$ \mu_{\rm total}$  &$ \mu_{\rm Ga}$ 
 &$ \mu_{\rm C}$  &$ \mu_{\rm Mn}$ \\
\tableline
$\rm PM $ &\dec 10.23 &\dec 0.17 &\dec 0.03 &\dec 3.34 
              &\dec . &\dec . &\dec . &\dec . \\
$\rm FM   $ &\dec 5.86 &\dec 0.16 &\dec 0.06 &\dec 1.88 
              &\dec 4.69 &\dec $-$0.12 &\dec $-$0.12 &\dec 1.64 \\
$\rm AFM   $ &\dec 5.36 &\dec 0.17 &\dec 0.05 &\dec 1.71 
              &\dec 0.00 &\dec 0.00 &\dec 0.00 &\dec 1.79 \\
\end{tabular}
\label{dos-mag}
\end{table}
  
\begin{figure}[tb]
\epsfig{file=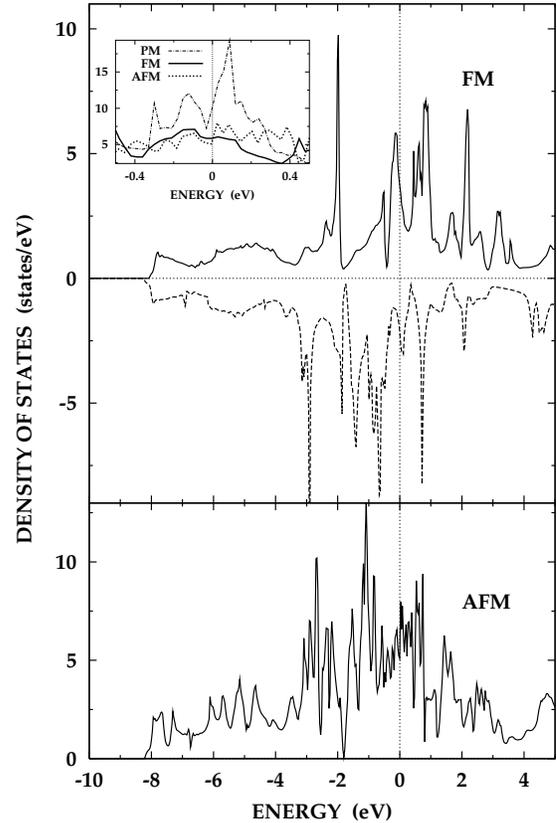,width=7.5cm}
\caption{Total DOSs of FM and AFM ground states of GaCMn$_3$. 
Inset compares the total DOSs of each magnetic
ground states including the PM ground state near Fermi level.}
\label{dos_mag}
\end{figure}

%
We have also investigated electronic structures of GaCMn$_3$ 
both in the FM and AFM phases.
Figure \ref{dos_mag} shows the total DOSs of the FM and AFM phases GaCMn$_3$. 
In the inset of Fig. \ref{dos_mag},
total DOSs near $E_{\rm F}$ in both phases are compared.
The high DOS in the PM phase is suppressed in the FM and AFM phases,
which have 5.86 and 5.36 (states/eV) at $E_{\rm F}$, respectively 
(Table \ref{dos-mag}).
It suggests that the FM and AFM phases are competing for the ground state.
Total energy calculation indicates that the FM phase has
lower energy than the AFM phase by $\sim$ 70 meV/GaCMn$_3$.
Thus it does not seem to be consistent with the experimental result 
in which the AFM phase is the ground state, but such a small energy difference
indicates that two phases are really competing and even the stable phase
would be susceptible to external conditions such as 
temperature, pressure, magnetic field, and C stoichiometry.

\begin{figure}[tb]
\epsfig{file=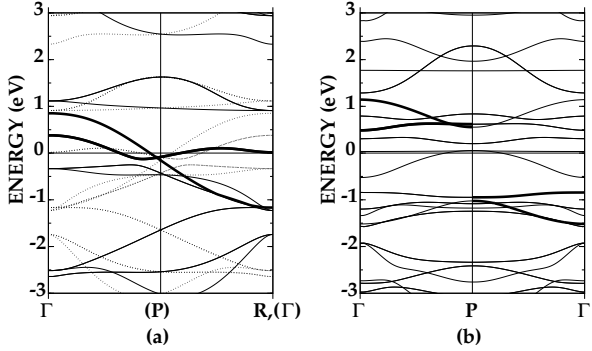,angle=270,width=7.5cm}
\caption{(a) The band structure of GaCMn$_3$
along the [111] direction in the PM phase. 
The dotted lines represent the bands folded around the 
point P when the simple-cubic unit cell is doubled along the
[111] direction. The thick solid lines correspond to the bands 
crossing the Fermi level.  
(b) The band structure of GaCMn$_3$ in the AFM phase 
along the line equivalent to (a).
Note the bands of thick solid lines which are split by about 1.5 eV. 
}
\label{band}
\end{figure}

To explore the effect of unit cell doubling in the AFM phase, the
band dispersion along the [111] direction is plotted in Fig. \ref{band}. 
For the comparison, the PM band structure calculated in the
same doubled unit cell is also provided together (Fig. \ref{band}(a)).
It is seen that the PM bands in the original sc unit cell
are folded around the point P which corresponds to $\frac{1}{2}$R in the 
simple-cubic 
unit cell. Noteworthy is that two bands crossing the Fermi level
along $\Gamma  {\rm R}$ in the PM phase (represented by thick solid lines) 
are split by $\sim 1.5$ eV due to the unit cell doubling in the AFM phase.
Thus, by band splitting, the DOS at $E_{\rm F}$ is reduced, and
accordingly the  AFM phase has effectively lower energy,
which indeed is consistent with the nesting scenario mentioned above.

Table \ref{dos-mag} provides magnetic moments in each magnetic phase.
For the  FM phase, the magnitude of Mn moment $\mu_{\rm Mn}$ is 
1.64 $\mu_{\rm B}$, whereas, for the AFM phase, 
$\mu_{\rm Mn}= 1.79\ \mu_{\rm B}$.
The magnetic moment for the AFM phase is in good agreement 
with experiment\cite{fruchart78} $\mu_{\rm Mn} = 1.8\ \mu_{\rm B}$ at 4.2 K.
As for the FM phase, $\mu_{\rm Mn}= 1.64\ \mu_{\rm B}$ seems to be
overestimated as compared to the experimental result $\mu_{\rm Mn} = 
1.3\ \mu_{\rm B}$ at 193 K. However, extrapolating the experimental 
value down to zero temperature would yield the magnetic moment 
close to the calculated one.
To check the contribution of the orbital polarization to the magnetic moment,
the LSDA+SO calculation including the spin-orbit interaction has been
performed. We have found the negligible orbital 
magnetic moments $\sim 0.04\ \mu_{\rm B}/{\rm Mn}$ in both phases, which
is consistent with the itinerant nature of GaCMn$_3$.

%
In conclusion,
we have performed the first principle electronic structure calculations 
on GaCMn$_3$ and related antiperovskite Mn compounds, 
SnCMn$_3$, ZnCMn$_3$, and ZnNMn$_3$.
In the PM phase, GaCMn$_3$
shows the Fermi surface nesting with wave vector 
${\rm\bf Q}\sim\Gamma{\rm R}$.
The nesting feature is confirmed by the calculated susceptibility.
Various magnetic structures of GaCMn$_3$ and related compounds
are well explained by the wave vectors which give rise to the maximum 
susceptibilities.
We have found that the DOS at $E_{\rm F}$ of AFM GaCMn$_3$ is similar
to that of FM GaCMn$_3$ which results in the competing ground 
state phase.
We have also demonstrated that the AFM GaCMn$_3$ is stabilized 
through the band folding effect.
The orbital moment of Mn is found to be negligible, as expected in
the itinerant magnetic system.

\acknowledgments
This work was supported by the KOSEF through the eSSC at POSTECH
and in part by the BK21 Project. Helpful discussions with N.H. Hur and
W.S. Kim are greatly appreciated.


\begin{thebibliography}{99}
%
\bibitem{kamishima00} K. Kamishima, T. Goto, H. Nakagawa, N. Minura,
                      M. Ohashi, N. Mori, T. Sasaki, and T. Kanomara,
\prb {\bf 63}, 024426 (2000).
%
\bibitem{kim} W. S. Kim, E. O. Chi, J. C. Kim, H. S. Choi, and  N. H. Hur,
Solid State Comm. {\bf 119}, 507 (2001).
%
\bibitem{he}  T. He, Q. Huang, A. P. Ramirez, Y. Wang, K. A. Reran,
              N. Rogado, M. A. Hayward, M. K. Haas, J. S. Slusky,
              Inumara, H. W. Zandbergen, N. P. Ong and R. J. Cava,
\nat {\bf 411}, 54 (2001).
%
\bibitem{fruchart70} D. Fruchart, E. F. Bertaut, F. Sayetat, M. Nasr Eddine,
                     R. Fruchart, and J. P. Senateur,
Solid State Comm. {\bf 8}, 91 (1970).
%
\bibitem{fruchart78} D. Fruchart and E. F. Bertaut,
J. Phys. Soc. Jap. {\bf 44}, 781 (1978).
%
\bibitem{kaneko} T. Kaneko, T. Kanomata, and K. Shirakawa,
J. Phys. Soc. Jap. {\bf 56}, 4047 (1987).
%
\bibitem{fruchart73} D. Fruchart, E. F. Bertaut, B. Le Clerc, Le Dang Khoi,
                     P. Veillet, G. Lorthioir, E. Fruchart, and R. Fruchart,
J. Solid State Chem. {\bf 44}, 781 (1973).
%
\bibitem{kamishima98} K. Kamishima, M. I. Bartashevich, T. Goto, M. Kikuchi,
                      and T. Kanomata,
J. Phys. Soc. Jap. {\bf 67}, 1748 (1998);
                     K. Kamishima, T. Goto, T. Kanomata, and M. I. Bartashevich,
J. Magn. Magn. Mater. {\bf 177-181}, 587 (1998);
                     K. Kamishima, T. Goto, and T. Kanomata,
Physica B {\bf 237}, 561 (1997).
%
\bibitem{kim-new} W. S. Kim, N. H. Hur,  {\it et al.}, private communications.
%
\bibitem{motizuki} K. Motizuki and H. Nagai,
J. Phys. C: Solid State Phys. {\bf 21} 5251 (1988).
%
\bibitem{myron} H. W. Myron, J. Rath, and A. J. Freeman,
\prb {\bf 15}, 885 (1977).
In our calculation, we have used an approximation of
the constant oscillator strength matrix element M$_{ij}$.
%
%
\end{thebibliography}
\end{document}